\def\lessthanorabout
  {\mathrel{\raise.3ex\hbox{$<$\kern-.75em\lower1ex\hbox{$\sim$}}}}
\def\greaterthanorabout
  {\mathrel{\raise.3ex\hbox{$>$\kern-.75em\lower1ex\hbox{$\sim$}}}} 

\def\today{\ifcase\month\or
  January\or February\or March\or April\or May\or June\or
  July\or August\or September\or October\or November\or December\fi
  \space\number\day, \number\year}

\def\boldr{{\bf r}}
\def\boldrzero{{\bf r_0}}
\def\scriptD{{\cal D}}
\def\scriptL{{\cal L}}

\def\setpsi{\{\Psi_n\}}

\documentstyle[aps,prb,psfig]{revtex}
\begin{document}
\bibliographystyle{jcp}
\title{Theory for the nonequilibrium dynamics of flexible chain molecules:
       relaxation to equilibrium of pentadecane from an all-trans
       conformation.}
\author{Wilfred H. Tang $\mbox{}^{1}$ \and
        Konstantin S. Kostov $\mbox{}^{2}$ \and
        Karl F. Freed $\mbox{}^{1}$ \\
$\mbox{}^{1}$James Franck Institute and the Department of Chemistry\\
$\mbox{}^{2}$James Franck Institute and the Department of Biochemistry
                and Molecular Biology \\
University of Chicago\\
Chicago, IL 60637\\
\today\\
}
\maketitle
\begin{abstract}
We extend to nonequilibrium processes our
recent theory for the long time dynamics of flexible chain molecules.
While the previous theory describes the equilibrium motions for any
bond or interatomic separation in (bio)polymers by time correlation
functions, the present extension of the theory enables the prediction
of the nonequilibrium relaxation
that occurs in processes, such as T-jump experiments, where there are
sudden transitions between, for example, different equilibrium states.
As a test of the theory, we consider the ``unfolding''
of pentadecane when it is transported from a constrained all-trans
conformation to a random-coil state at thermal equilibrium.
The time evolution of the mean-square end-to-end distance
$\langle R_{\rm end}^2(t) \rangle_{\rm noneq}$
after release of the constraint is computed
both from the theory and from Brownian dynamics (BD) simulations. The lack of
time translational symmetry for nonequilibrium processes requires that
the BD simulations of the relaxation of
$\langle R_{\rm end}^2(t) \rangle_{\rm noneq}$
be computed from an average over
a {\it huge number} of independent trajectories, rather than over
successive configurations from a single trajectory, which may be used to
generate equilibrium time correlation functions.
Adequate convergence ensues for the non-equilibrium simulations
only after averaging 9000 trajectories, each of 0.8 ns duration.
In contrast, the theory
requires only equilibrium averages for the initial and final
states, which may be readily obtained from a {\it few} Brownian dynamics
trajectories.
Therefore, the new method produces enormous savings in computer time.
Moreover, since both theory and simulations use identical potentials and
solvent models the theory contains no adjustable parameters. The predictions
of the theory for the relaxation of
$\langle R_{\rm end}^2(t) \rangle_{\rm noneq}$
agree very well
with the BD simulations.
This work is a starting point for the application of
the new method to nonequilibrium processes with biological importance such
as the helix-coil transition and protein folding.
\end{abstract}

\section{Introduction}
The nonequilibrium dynamics of polymers is important for understanding a
variety of phenomena.  For instance, the dynamics of protein
folding/unfolding is a fundamental biological process.
Our understanding of this process , however, is severely limited by
a shortage of adequate theoretical methods for studying
the protein {\it dynamics} with realistic molecular models.
Experimentally, it is difficult
to achieve sufficiently short time resolution to follow the full dynamics of
protein folding/unfolding.  Conversely, the main theoretical technique,
molecular dynamics, cannot reach time scales long enough to probe many of
the interesting events due to the tremendous amount of computation required.

We have been developing a theory for the long time dynamics of flexible
polymers and polypeptides in solution.  Our mode coupling
theory\cite{chang1,chang2,tang,kostov1,kostov2} is based on a 
significant extension of the generalized Rouse-Zimm
theory\cite{orzzwanzig,bixonzwanzig,pericoreview,perico1,perico2,perico3,perico4,perico5,yi1,yi2}
to include contributions to the dynamics from the memory function
(``internal friction'') terms that are customarily ignored.
Prior papers demonstrate that the mode coupling theory produces an
excellent representation of the long time equilibrium time correlation
functions for both alkanes and short polypeptides.
The present paper describes the extension of the mode coupling theory to
treat the nonequilibrium dynamics occurring when a system in equilibrium
(or in a constrained equilibrium) is suddenly transported to a different
equilibrium (perhaps constrained) state, such as in T-jump experiments.

We apply the theory to describe the relaxation of pentadecane from
nonequilibrium initial conditions to equilibrium at a temperature of 300 K.
The pentadecane chain begins in its maximally stretched conformation,
in which all the dihedral angles are trans.  The molecule subsequently
relaxes to equilibrium, where there is a distribution of dihedral angles
due to thermal fluctuations.
The direct simulation of this non-equilibrium process requires orders of
magnitude more computer time than is necessary for simulating equilibrium
time correlation functions.  The complexity of the non-equilibrium
simulations is the reason pentadecane is taken to begin in the all-trans
configuration, rather than using an initial thermal distribution at a 
temperature lower than 300 K, where a huge number of additional
non-equilibrium trajectories would be necessary to average over the thermal
distribution of initial conditions.
In contrast, the treatment of such a T-jump process by the theory
merely doubles the more limited computational effort
necessary to describe the dynamics beginning from the all-trans pentadecane.
Studying the ``unfolding'' dynamics of this
relatively simple system provides insight into how to apply
the mode coupling theory to more complicated biological systems,
such as the unwinding of an $\alpha$-helix or the unfolding of a protein.

A crucial step in the theory is the selection of a suitable basis set.
Due to symmetry, the basis functions for describing nonequilibrium dynamics
must differ from those previously used for calculating
equilibrium dipole time correlation functions.
However, the same principles apply independent of
the dynamical observables of interest.  For equilibrium dipole time
correlation functions, a simple first-order basis set produces reasonable
estimates\cite{chang1,chang2,tang,kostov1,kostov2} of alkane dynamics,
and a more sophisticated
sorting method yields excellent approximations while maintaining a
manageable basis set size.\cite{kostov1,kostov2}
The present paper extends our mode coupling procedure to determine the
appropriate basis sets for describing the long time portions of the
nonequilibrium dynamics.
As in previous studies, we test the effectiveness of various basis sets
by comparing the predictions of the theory with ``exact'' results from
Brownian dynamics simulations.  Because all the equilibrium inputs required
by the theory are taken from the Brownian dynamics simulations,
there are {\it no} free parameters in this comparison.

Section II presents the general theory for describing the nonequilibrium
dynamics, while the BD simulation methods and united atom alkane
potentials are described in the subsequent two sections.
Theory and simulations are compared in Section V with {\it no} adjustable
parameters.  The final section discusses the potential application of
the theory to more complex systems, such as proteins.

\section{Nonequilibrium averages from the theory}
We consider a polymer consisting of $N$ beads.  Denote the $3N$
Cartesian coordinates of the polymer beads by
$\boldr(t) = \{ r_1(t), r_2(t), \ldots, r_{3N}(t) \}
           = \{ x_1(t), y_1(t), z_1(t), x_2(t), y_2(t), z_2(t), \ldots,
                x_N(t), y_N(t), z_N(t) \}                                 $.
The dynamics of this model polymer are assumed to be governed
by the Smoluchowski equation,\cite{doiedwards,gardiner,risken}
\begin{equation}
{{\partial P(\boldr,t \mid \boldrzero,0)} \over {\partial t}} =
  \scriptD(\boldr) P(\boldr,t \mid \boldrzero,0)  ,
\label{eq:smoluchowski}
\end{equation}
\begin{equation}
\scriptD(\boldr) =
\sum_{i=1}^{3N}
\left( {{k_B T}
\over {\zeta}}
{{\partial^2} \over {\partial r_i^2}} + 
{{1} \over {\zeta}}
{{\partial} \over {\partial r_i}}
{{\partial U(\boldr)} \over {\partial r_i}}
\right)  ,
\label{eq:D}
\end{equation}
where
\hbox{$P(\boldr,t \mid \boldrzero,0)$} denotes the probability that the
coordinates of the $N$ beads at time $t$ are $\boldr$,
given that the coordinates of
the $N$ beads at time $0$ are $\boldrzero$,
$k_B$ is Boltzmann's constant, $T$ is the absolute temperature,
$U(\boldr)$ is the potential energy,
and $\zeta$ is the bead friction coefficient.
The only explicit degrees of freedom in these equations are the positions
$\boldr$ of the polymer beads.  The solvent affects the motion of the polymer
through implicit frictional and stochastic forces.
The frictional force is taken to be
large enough that the motion lies is in the overdamped regime --- that is, 
inertial effects may be neglected.
We also neglect hydrodynamic interactions between different beads, but the
general theory below may readily be applied including hydrodynamic
interactions in equation~(\ref{eq:D}).
The hydrodynamic interactions are omitted here to reduce the computational
labor but have been treated previously within the generalized Rouse-Zimm
theory.\cite{yi1}
The simplifying nature of these assumptions is useful for specifying a model
for which it is possible to make rigorous, parameter-free comparisons
between theory and ``exact'' results from computer simulations of
equation~(\ref{eq:smoluchowski}).
Moreover, this particular model contains many of the important
physical features without being overly complicated.
After understanding the application of the theory to this relatively simple
system, the theory can be generalized to treat more realistic and complex
systems.  For example, the equilibrium theory has recently been applied to
describe the dynamics in alkane melts where the dynamics is governed by
the full Liouville equation.\cite{kostov2}

We are interested in describing the relaxation of a polymer from
nonequilibrium starting conditions to equilibrium.
In the example illustrated below, the nonequilibrium initial condition
is the all-trans conformation of pentadecane, while in the final state,
the pentadecane chain presents a large distribution of conformations.
Denote by $\langle f(t) \rangle_{\rm noneq}$
the expected value of some quantity $f(t)$ for a system that starts from
nonequilibrium starting conditions at time $t=0$.
The nonequilibrium average $\langle f(t) \rangle_{\rm noneq}$
can be expressed in terms of probability distributions as
\begin{equation}
\langle f(t) \rangle_{\rm noneq}
 = \int d\boldrzero P(\boldrzero,0)
   \int d\boldr f(\boldr) P(\boldr,t \mid \boldrzero,0)  ,
\label{eq:f1}
\end{equation}
where $P(\boldrzero,0)$ denotes the probability that the coordinates of the
$N$ polymer beads at time $0$ are $\boldrzero$,
\hbox{$P(\boldr,t \mid \boldrzero,0)$} is the formal solution to
equation~(\ref{eq:smoluchowski}),
\begin{equation}
P(\boldr,t \mid \boldrzero,0)
 = e^{t \scriptD(\boldr)} P(\boldr,0 \mid \boldrzero,0)
 = e^{t \scriptD(\boldr)} \delta (\boldr - \boldrzero)  ,
\label{eq:P}
\end{equation}
and $\delta(\boldr - \boldrzero)$ is the Dirac delta function.
Thus, equation~(\ref{eq:f1}) may be rewritten as
\begin{eqnarray}
\langle f(t) \rangle_{\rm noneq}
 &=& \int d\boldrzero P(\boldrzero,0)
     \int d\boldr f(\boldr) e^{t \scriptD(\boldr)} \delta(\boldr - \boldrzero)
\nonumber
\\
 &=& \int d\boldrzero P(\boldrzero,0)
     \int d\boldr \delta(\boldr - \boldrzero) e^{t \scriptL(\boldr)} f(\boldr)
\nonumber
\\
 &=& \int d\boldrzero P(\boldrzero,0)
     e^{t \scriptL(\boldrzero)} f(\boldrzero)  ,
\label{eq:f2}
\end{eqnarray}
where $\scriptL$ is the adjoint of $\scriptD$,
\begin{equation}
\scriptL(\boldr)
 = \scriptD^{\dag}(\boldr)
 = \sum_{j=1}^{3N} \left( {{k_B T} \over {\zeta}}
   {{\partial^2} \over {\partial r_j^2}} - 
   {{1} \over {\zeta}} {{\partial U(\boldr)} \over {\partial r_j}}
   {{\partial} \over {\partial r_j}} \right).
\end{equation}

The formal expression in equation~(\ref{eq:f2}) is rendered computationally
tractable by introducing the eigenfunctions $\Psi_n$ and
eigenvalues $\lambda_n$ of $\scriptL$,
\begin{equation}
\scriptL \Psi_n = -\lambda_n\Psi_n, \qquad n=1,2,\ldots . 
\label{eq:eigenvalueL}
\end{equation}
Because $e^{-U/(k_B T)} \scriptL$ is self-adjoint, 
the natural inner product is defined by
\begin{equation}
(\chi_i,\chi_j) 
  = {{\int d \boldr \exp[-U(\boldr)/(k_B T)]\chi_i \chi_j}
  \over {\int d \boldr \exp[-U(\boldr)/(k_B T)]}}
\end{equation}
and coincides with the definition of the {\it equilibrium} average,
\begin{equation}
\langle \chi_i \chi_j \rangle
  = {{\int d \boldr \exp[-U(\boldr)/(k_B T)]\chi_i \chi_j}
  \over {\int d \boldr \exp[-U(\boldr)/(k_B T)]}}  .
\end{equation}
With this choice of inner product, a complete, orthonormal set $\setpsi$
of eigenfunctions can, in principle, be found.
Thus, the $f(\boldrzero)$ of equation~(\ref{eq:f2}) can be expanded
in terms of the eigenfunctions $\setpsi$ as
\begin{equation}
f(\boldrzero)
 = \sum_n (f, \Psi_n) \Psi_n(\boldrzero)
 = \sum_n \langle f \Psi_n \rangle \Psi_n(\boldrzero) .
\label{eq:expansion}
\end{equation}
Substituting equation~(\ref{eq:expansion}) into equation~(\ref{eq:f2}) produces
\begin{equation}
\langle f(t) \rangle_{\rm noneq}
 = \int d\boldrzero P(\boldrzero,0)
   e^{t \scriptL(\boldrzero)}
   \sum_n \langle f \Psi_n \rangle \Psi_n(\boldrzero)
 = \sum_n e^{-t \lambda_n} \langle f \Psi_n \rangle
   \int d\boldrzero P(\boldrzero,0) \Psi_n(\boldrzero)
\label{eq:f3}
\end{equation}

Since the eigenvalue equation~(\ref{eq:eigenvalueL}) cannot be solved
analytically for any realistic potential, the solutions are approximated
using a basis set expansion,
\begin{equation}
\Psi_n \approx \sum_{i=1}^{M}C_{in}\phi_i.
\label{eq:psi_n_approximate}
\end{equation}
The coefficients $C_{in}$ are therefore
determined by solving the matrix eigenvalue
problem,\cite{lapack}
\begin{equation}
{\bf F C}={\bf S C \Lambda},
\label{eq:matrixeigenvalue}
\end{equation}
subject to the normalization constraint,
\begin{equation}
{\bf C}^T {\bf S C} = {\bf I},
\end{equation}
where ${\bf I}$ is the unit matrix,
\begin{equation}
S_{ij} = \langle \phi_i \phi_j \rangle,
\end{equation}
\begin{equation}
F_{ij} = \langle \phi_i \scriptL \phi_j \rangle
       = \sum_{k=1}^{3N}
           { -{k_B T \over \zeta}
              \langle {{\partial {\phi_i}} \over {\partial {r_k}}}
                      {{\partial {\phi_j}} \over {\partial {r_k}}} \rangle
           },
\label{eq:F_ij}
\end{equation}
and ${\bf \Lambda}$ is the diagonal matrix of (approximate) eigenvalues
$\lambda_1', \lambda_2',\ldots, \lambda_M'$.
Substituting the $C_{in}$ and $\lambda_n'$ obtained from
the matrix eigenvalue problem into equation~(\ref{eq:f3}) produces the
final equation for $\langle f(t) \rangle_{\rm noneq}$ as
\begin{equation}
\langle f(t) \rangle_{\rm noneq} \approx
  \sum_{n=1}^M \left( \sum_{i=1}^M C_{in} \langle f \phi_i \rangle \right)
               \left( \sum_{j=1}^M C_{jn}
                      \int d\boldrzero P(\boldrzero,0) \phi_j(\boldrzero)
               \right)
               e^{-\lambda_n' t}.
\label{eq:finalresult}
\end{equation}
The nonequilibrium average in equation~(\ref{eq:finalresult}) is
evaluated using as input only equilibrium information
($\langle f \phi_i \rangle$, ${\bf F}$, and ${\bf S}$)
at the final temperature $T$ and similar averages
($\int d\boldrzero P(\boldrzero,0) \phi_j(\boldrzero)$)
for the initial condition.
In our case, the initial condition is the all-trans conformation of
pentadecane, and ``averaging'' over the initial condition simply involves
evaluating $\phi_j$ for the all-trans conformation.
However, equation~(\ref{eq:finalresult}) can be used for more general
nonequilibrium starting conditions, such as an initial temperature
$T_{\rm init}$ (or pressure) different from the final temperature $T$,
in which case $\int d\boldrzero P(\boldrzero,0) \phi_j(\boldrzero)$ is
calculated by averaging $\phi_j$ over the Boltzmann distribution
corresponding to $T_{\rm init}$.
In fact, the initial condition could even correspond to a state
(equilibrium or constrained) of the polymer in a different solvent.

\section{Brownian dynamics simulations}
The BD computer simulations are more conveniently pursued by rewriting
the Smoluchowski equation~(\ref{eq:smoluchowski}) as a set of equivalent
Langevin equations,
\begin{equation}
{dr_i(t) \over dt} = -{1 \over \zeta} {\partial U[\boldr(t)]
\over \partial r_i(t)} + X_i^*(t), \qquad i=1,2,\ldots,3N,
\end{equation}
where $X_i^*(t)$ is a Gaussian random variable
with the  properties,
\begin{displaymath}
\langle X_i^*(t) \rangle = 0,
\end{displaymath}
\begin{equation}
\langle X_i^*(t) r_j(t') \rangle = 0, \qquad \hbox{for } t > t',
\end{equation}
\begin{displaymath}
\langle X_i^*(t) X_j^*(t') \rangle = 2{k_B T \over \zeta}
\delta_{ij} \delta(t-t'),
\end{displaymath}
where $\delta_{ij}$ is the Kronecker delta function
and $\delta(t-t')$ is the Dirac delta function.
The Brownian dynamics simulations are performed using the
algorithm of van Gunsteren and Berendsen,\cite{brownian,allentildesley}
\begin{equation}
r_i(n+1) = r_i(n) + {F_i(n) \over \zeta} \Delta t
           + {1 \over 2} {F_i'(n) \over \zeta} (\Delta t)^2 + \chi_{in},
\qquad i=1,\ldots,3N,
\end{equation}
where $\Delta t$ is the time step,
$F_i(n) = - \partial U({\bf r}) / \partial r_i$ is the force acting on the
$i$th coordinate, $F_i'(n) = [F_i(n) - F_i(n-1)] / \Delta t$ is the finite
difference approximation to the time derivative of the force, and
$\chi_{in}$ is a random positional displacement taken from a Gaussian
distribution with zero mean and variance $2 (k_B T / \zeta) \Delta t$.

The calculation of the nonequilibrium averages
$\langle f(t) \rangle_{\rm noneq}$ from the BD simulations uses
a series of trajectories run at a temperature of 300 K starting
with pentadecane in the all-trans conformation where the
molecule lies in the $x-y$ plane with the long axis parallel to the
$x$ axis and with the first three beads having the coordinates
$( 0.00 {\rm \AA}, 0.87 {\rm \AA}, 0.00 {\rm \AA} ),$
$( 1.26 {\rm \AA}, 0.00 {\rm \AA}, 0.00 {\rm \AA} ),$ and
$( 2.52 {\rm \AA}, 0.87 {\rm \AA}, 0.00 {\rm \AA} )$.
Because all of the trajectories differ greatly from one another,
obtaining good averages requires many trajectories,
and the simulations use 9000 trajectories, each of 800 ps duration.

The theory of section III calculates the equilibrium averages
$\langle \cdots \rangle$ from a single
400 ns trajectory run at a temperature of 300 K.  The system is allowed
to equilibrate for 20 ns, after which data are collected for 380 ns.

\section{Polymer model}
Pentadecane is modeled as a string of 15 ${\rm CH}_2$ or ${\rm CH}_3$
united atom groups.  The potential $U$ is, except for one modification, that
of the GROMOS package.\cite{gromos}
This potential has the general form
\begin{equation}
U = U_{\rm bond} + U_{\rm angle} + U_{\rm dihedral} + U_{\rm nonbond}.
\end{equation}
The bond length potential is
\begin{equation}
U_{\rm bond} = \sum_{i=1}^{14} {1 \over 2} K_b (l_i - l_0)^2,
\end{equation}
where $l_i$ is the length of bond $i$,
$l_0 = 1.53 \; {\rm \AA}$ is the GROMOS equilibrium bond length for alkanes,
and $K_b$ is the bond length force constant.
We use the force constant
$K_b = 160 \; {\rm kcal} \; {\rm mol}^{-1} {\rm \AA}^{-2}$,
which is smaller than
$K_b$ in GROMOS by a factor of 5.  This change does not significantly affect
the long-time dynamics but allows using a larger Brownian dynamics time step
($\Delta t = 5 \; {\rm fs}$).
The bond angle potential is
\begin{equation}
U_{\rm angle} = \sum_{i=1}^{13} {1 \over 2} K_{\theta}
                (\theta_i - \theta_0)^2,
\end{equation}
where $\theta_i$ is the $i$th bond angle,
$\theta_0 = 111^{\rm o}$ is the GROMOS equilibrium bond angle for alkanes,
and $K_\theta = 110 \; {\rm kcal} \; {\rm mol}^{-1} {\rm rad}^{-2}$ is the
bond angle force constant.
The dihedral angle potential is
\begin{equation}
U_{\rm dihedral} = \sum_{i=1}^{12} K_\phi (1 + \cos 3\phi_i),
\end{equation}
where $\phi_i$ is the $i$th dihedral angle,
and $K_\phi = 1.4 \; {\rm kcal} / {\rm mol}$.
The nonbonded potential $U_{\rm nonbond}$ uses a 6-12 Lennard-Jones
potential to describe the interaction between two nonbonded united atoms.
More details are given in Ref.~\onlinecite{gromos}.

The solvent is modeled as a structureless continuum that exerts a viscous
damping force on the pentadecane chain.
The friction coefficient is calculated using Stokes' law
$\zeta = 6 \pi \eta R$, where the viscosity $\eta$ is taken to be 1 cp
and the hydrodynamic radius of a ${\rm CH}_2$ or ${\rm CH}_3$ united atom
group is taken to be $1.5 \; {\rm \AA}$.

\section{Results}
The only approximation in the theory is associated with the use of a finite
basis set in equation~(\ref{eq:psi_n_approximate}).
Thus, we begin by studying the effectiveness of various basis sets for
estimating the time evolution of the mean-square end-to-end distance 
$R_{\rm end}^2 = (x_{15} - x_1)^2 + (y_{15} - y_1)^2 + (z_{15} - z_1)^2$.
In contrast to the previously studied equilibrium time correlation
functions of dynamical quantities that are odd in the position
coordinates,\cite{chang1,chang2,tang,kostov1,kostov2}
the mean-square
end-to-end distance is an even function of the position coordinates.
Therefore, symmetry considerations dictate the use of different
(even-power, as opposed to odd-power)
basis functions in the mode-coupling theory calculation of
$\langle R_{\rm end}^2(t) \rangle_{\rm noneq}$.
The scalar nature of $R_{\rm end}^2(t)$ implies that a general
mode coupling basis may be constructed from the basis functions\cite{footnote1}
\begin{displaymath}
1,
\end{displaymath}
\begin{displaymath}
{\bf l_i} \cdot {\bf l_j} \;
[\forall \mbox{ (distinct) pairs } \{i,j\}, i,j = 1, \ldots, 14],
\end{displaymath}
\begin{displaymath}
({\bf l_i} \cdot {\bf l_j}) ({\bf l_k} \cdot {\bf l_m}) \;
[\forall \mbox{ (distinct) pairs of (distinct) pairs }
\{ \{i,j\},\{k,m\} \}, i,j,k,m = 1, \ldots, 14],
\end{displaymath}
\begin{displaymath}
({\bf l_i} \cdot {\bf l_j}) ({\bf l_k} \cdot {\bf l_m})
({\bf l_n} \cdot {\bf l_p}) \;
[\forall \mbox{ (distinct) triples of (distinct) pairs, }
\end{displaymath}
\begin{displaymath}
\{ \{i,j\},\{k,m\},\{n,p\} \}, i,j,k,m,n,p = 1, \ldots, 14],
\end{displaymath}
\begin{displaymath}
\mbox{etc.},
\end{displaymath}
where ${\bf l_i}$ denotes the $i$th pentadecane bond vector with components
$l_{ix} = x_{i+1} - x_i$,
$l_{iy} = y_{i+1} - y_i$, and
$l_{iz} = z_{i+1} - z_i$,
for $i = 1, \ldots, 14$.
The ordering of pairs (and triples, etc.) is unimportant;
for example, the pair $\{1,2\}$ is equivalent to $\{2,1\}$ since
${\bf l_1} \cdot {\bf l_2} = {\bf l_2} \cdot {\bf l_1}$,
and hence only one of them is included in the basis set.
The constant basis function 1, which produces a vanishing eigenvalue in
equation~(\ref{eq:matrixeigenvalue}), accounts for the non-zero value of
$\langle R_{\rm end}^2(t) \rangle_{\rm noneq}$
at infinite times.

The mode coupling theory is first applied using the relatively simple basis
set,
\begin{displaymath}
\{ 1, \;
   {\bf l_i} \cdot {\bf l_j} \;
   [\forall \mbox{ (distinct) pairs } \{i,j\}, i,j = 1, \ldots, 14] \} ,
\qquad \mbox{(Basis set I)}.
\end{displaymath}
Fig.~1 demonstrates that basis set I provides a reasonable estimate
for the time evolution of
$\langle R_{\rm end}^2(t) \rangle_{\rm noneq}$
whose numerically ``exact'' dynamics is represented by the BD simulation.
A more accurate approximation of
$\langle R_{\rm end}^2(t) \rangle_{\rm noneq}$
requires expanding basis set I to include more basis functions
of the appropriate symmetry, such as the tetralinear products
$({\bf l_i} \cdot {\bf l_j}) ({\bf l_k} \cdot {\bf l_m})$.
However, the total number of tetralinear products scales as $N^4$,
where $N$ is the number of monomers.
Thus, inclusion of all tetralinear functions makes the basis set unmanageably
large even for the relatively small ``polymer'' pentadecane.
Clearly, all the tetralinear products cannot generally be retained in the
basis set, nor are all these basis functions relevant to the long time
dynamics.
Those tetralinear products that contribute the most to
the long time dynamics are determined using a selection technique similar
to that introduced previously for computing equilibrium time correlation
functions.\cite{tang,kostov1,kostov2}

Based on prior treatments of equilibrium time correlation functions,
it proves convenient to work with basis functions constructed as
linear combinations of individual bond vectors that provide a first
approximation to the collective chain motions.
An approximate set of  eigenfunctions, denoted as the first order
GR (generalized Rouse) modes, of $\scriptL$ may be obtained from
equations~(\ref{eq:psi_n_approximate}) - (\ref{eq:F_ij}) with a basis set
consisting of all linear functions in the bond vectors,
\begin{displaymath}
\{ l_{ix} [i = 1, \ldots, 14],
   l_{iy} [i = 1, \ldots, 14],
   l_{iz} [i = 1, \ldots, 14] \}.
\end{displaymath}
The first order GR modes are thus linear combinations of the bond vector
components.
Note that the linear basis functions and the corresponding first order
GR modes do not contribute to the expression in equation~(\ref{eq:finalresult})
for $\langle R_{\rm end}^2(t) \rangle_{\rm noneq}$
due to symmetry considerations.
Products of first order GR modes, however, may be used as providing a first
approximation to $\langle R_{\rm end}^2(t) \rangle_{\rm noneq}$.
Due to the invariance of $\scriptL$ under interchange of the Cartesian
directions $x$, $y$, and $z$, the first order GR modes need only be
computed for one direction, say $x$, and the GR modes in the other two
directions follow trivially by symmetry.
Let $\Psi_{1x}^{(1)}, \Psi_{2x}^{(1)}, \ldots, \Psi_{14x}^{(1)}$
be the first order GR modes obtained from solving
equations~(\ref{eq:psi_n_approximate}) - (\ref{eq:F_ij}),
and let their corresponding (approximate) eigenvalues 
$\lambda_1^{(1)}, \lambda_2^{(1)}, \ldots, \lambda_{14}^{(1)}$
be ordered in increasing magnitude.
Thus, the first eigenfunction $\Psi_{1x}^{(1)}$ provides
the first order estimate to the GR
eigenfunction with the slowest decay, called the slowest mode,
$\Psi_{2x}^{(1)}$ provides the first order estimate to the GR eigenfunction of
the next slowest decaying mode, and so on.
Let $\Psi_{1y}^{(1)}, \Psi_{2y}^{(1)}, \ldots, \Psi_{14y}^{(1)}$
be the GR modes obtained by replacing all the $x$'s in
$\Psi_{1x}^{(1)}, \Psi_{2x}^{(1)}, \ldots, \Psi_{(N-1)x}^{(1)}$ with $y$'s, and
let $\Psi_{1z}^{(1)}, \Psi_{2z}^{(1)}, \ldots, \Psi_{14z}^{(1)}$
be the corresponding GR modes in $z$.
It is convenient to introduce the notation
${\bf \Psi_i^{(1)}} [i=1,2,\ldots,14]$ to represent the vector
$( \Psi_{ix}^{(1)}, \Psi_{iy}^{(1)}, \Psi_{iz}^{(1)} )$.
We expect that the slowest decaying first order eigenfunctions contain the
most information about the long-time dynamics.
Thus, rather than including all possible tetralinear products in the
basis set, we only include tetralinear products constructed
from the slowest decaying GR modes.
This produces the basis set which contains the functions\cite{footnote2}
\begin{displaymath}
\{ 1, \;
   {\bf \Psi_i^{(1)}} \cdot {\bf \Psi_j^{(1)}} \;
   [\forall \mbox{ (distinct) pairs } \{i,j\}, i,j = 1, \ldots, 14],
\end{displaymath}
\begin{displaymath}
   ({\bf \Psi_i^{(1)}} \cdot {\bf \Psi_j^{(1)}})
   ({\bf \Psi_k^{(1)}} \cdot {\bf \Psi_m^{(1)}}) \;
   [\forall \mbox{ (distinct) pairs of (distinct) pairs }
   \{ \{i,j\},\{k,m\} \}, i, j = 1, \ldots, Q] \} ,
\end{displaymath}
\begin{displaymath}
\mbox{(Basis set II}_Q),
\end{displaymath}
where $0 \le Q \le 14$.
The $Q=0$ case reduces basis set ${\rm II}_{Q=0}$ to basis set I,
while using $Q=14$ is equivalent to including all possible tetralinear
products in the basis set.  Note that the size of basis set ${\rm II}_Q$
scales roughly as $N^2+Q^4$.  Hence, basis set ${\rm II}_Q$ is
significantly smaller than the full basis set containing all
possible tetralinear products if $Q \ll N$.
Fig.~1 demonstrates that basis set ${\rm II}_Q$ with $Q=4$ produces an
improvement in the long time dynamics predicted by the theory,
though for short times below 100 ps, basis set I is in better
agreement with the simulations.
Increasing $Q$ beyond 4 does not produce any significant improvement.
Greater accuracy of the theory therefore requires the inclusion of hexalinear
and higher order products in the basis set.
However, the size of the basis set grows extremely rapidly as higher
order functions are included, even when a procedure similar to
that in basis set ${\rm II}_Q$ is used.
In order to select those higher order that contribute the most to
the long time dynamics, we now test an alternative sorting procedure
first described in Ref.~\onlinecite{kostov1} for treating the long
time dynamics of small flexible peptides.

The sorting procedure orders basis functions according to the sum
of the (approximate) eigenvalues of the GR modes in the tetralinear and
higher order (even) products of GR modes.
For example,
$\lambda_i^{(1)} + \lambda_j^{(1)} + \lambda_k^{(1)} + \lambda_m^{(1)}$
represents the first order relaxation rate for the tetralinear product
$({\bf \Psi_i^{(1)}} \cdot {\bf \Psi_j^{(1)}})
 ({\bf \Psi_k^{(1)}} \cdot {\bf \Psi_m^{(1)}})$.
Since we expect that the functions most relevant to the long time dynamics are
composed of the slowest-decaying GR modes (smallest approximate eigenvalues),
the basis set retains the products with the smallest eigenvalue sums.
While first order eigenvalues may not provide an optimal estimate for the
relative importance of different products of GR modes,
previous experience suggests that this method is sufficient
for obtaining excellent equilibrium time correlation
functions.\cite{kostov1,kostov2}
The first order eigenvalue sorting procedure produces the basis set which
is represented as
\begin{displaymath}
\{ 1, \;
   {\bf \Psi_i^{(1)}} \cdot {\bf \Psi_j^{(1)}} \;
   [\forall \mbox{ (distinct) pairs } \{i,j\}, i,j = 1, \ldots, 14],
\end{displaymath}
\begin{displaymath}
\mbox{higher order GR mode products (up to 14 products) with the }R
\mbox{ smallest eigenvalue sums}
\}  ,
\end{displaymath}
\begin{displaymath}
\mbox{(Basis set III}_R).
\end{displaymath}
Fig.~1 indicates that basis set ${\rm III}_R$ with $R=500$
produces further improvement in the theory.
Increasing $R$ beyond 500 results in numerical instabilities in the
calculation, though previous experience indicates that using longer
equilibrium simulations rectifies the problem.\cite{kostov1,kostov2}

The discrepancy in the ``infinite'' time $R^2_{\rm end}$ in Fig.~1
is due to the difference between the ``infinite'' time (i.e., equilibrium)
$\langle R^2_{\rm end}(t) \rangle_{\rm noneq}$
from the 9000 nonequilibrium BD simulations and the
$\langle R^2_{\rm end} \rangle$ from the 400 ns equilibrium BD simulation.
This difference arises from statistical fluctuations in calculating
averages, and the magnitude of these fluctuations can be estimated by a
simple calculation.
The variation in $R^2_{\rm end}$ is comparable to the size of $R^2_{\rm end}$
itself --- about 2 ${\rm nm}^2$.
In a BD simulation,  the ``memory'' of previous values of $R^2_{\rm end}$ is
about 100 ps (see Fig.~1), so a 400 ns BD simulation yields about
4000 independent samples of $R^2_{\rm end}$.
Because $R^2_{\rm end}$ is {\it not} normally distributed, the
central limit theorem is not strictly applicable.\cite{statisticsbook}
Nevertheless, the theorem can provide an estimate
for the order of magnitude of the fluctuations in calculating the average
$\langle R^2_{\rm end} \rangle$ as
$2 \; {\rm nm}^2 / \sqrt{4000} \approx 0.03 \; {\rm nm}^2$.
This value is consistent with the discrepancy in Fig.~1 and
is also consistent with the variation in
the equilibrium average $\langle R^2_{\rm end} \rangle$
between two different 400 ns BD simulations (data not shown).

Fig.~2 demonstrates that the relaxation of $R^2_{\rm end}$
to equilibrium is highly nonexponential at long times.
This is in contrast to equilibrium dipole time correlation functions,
where the exponential decay at long times is predicted
very accurately by the mode coupling theory.\cite{kostov1}
The nonexponential behavior of
$\langle R^2_{\rm end}(t) \rangle_{\rm noneq}$ at long times may account
for the difficulty in obtaining highly accurate calculations from the theory.
Nevertheless, it is encouraging that the mode coupling theory gives
reasonable estimates for $\langle R^2_{\rm end}(t) \rangle_{\rm noneq}$, and,
furthermore, correctly predicts that the relaxation at long times is
nonexponential.

\section{Discussion}
The computational effort required to calculate
$\langle R_{\rm end}^2(t) \rangle_{\rm noneq}$
directly from Brownian dynamics simulations is enormous and
requires orders of magnitude more computer time than is necessary for
simulating equilibrium time correlation functions.
The absence of time translational symmetry during a nonequilibrium
relaxation precludes the computation of this process from a single
Brownian dynamics trajectory.  Instead, an ensemble average must be
performed over a {\it huge} number of nonequilibrium trajectories.
For example, 9000 trajectories are necessary to obtain with sufficiently
small noise the relaxation to equilibrium of
$\langle R_{\rm end}^2(t) \rangle_{\rm noneq}$ for pentadecane
from the all-trans conformation.
The accumulation of these nonequilibrium trajectories requires more than
one month on an SGI Power Challenge.
Moreover, this simulation is far simpler than that
needed for most nonequilibrium processes.
The direct simulation of a T-jump process would necessitate averaging over
the distribution of both initial and final conditions and would require
at least an order of magnitude more than the 9000 trajectories used for the
simulation with constrained initial conditions.
Clearly, any realistic study of the unfolding of a protein by
computer simulations is currently unfeasible and will remain so even in the
more distant future.  In contrast, the theory requires as inputs
only {\it equilibrium} averages which may be obtained readily from a few
Brownian dynamics trajectories.
Furthermore, the theory can easily handle many different types of
nonequilibrium processes.
Treating the relaxation of pentadecane from the all-trans conformation
requires only equilibrium averages for the final state.
To describe a T-jump process using the theory, equilibrium averages must
be computed for both the initial state and the final state --- merely
a doubling of the computational effort.

The successful description by the theory of the ``unfolding'' of pentadecane
indicates that the theory proposed here may provide a significant
contribution to the study of protein folding and unfolding.
Understanding how a globular protein folds requires detailed
characterization of partially organized intermediates formed during the
folding process.  Information about such intermediates is becoming
available through many complementary experimental studies involving
stable partially folded states, protection from hydrogen exchange,
disulfide formation in oxidative refolding, effects of amino acid
substitutions, and peptide analogues of folding intermediates.\cite{dobson}
Nonetheless, no clear unifying view of the nature of folding mechanisms
has yet emerged, and many uncertainties over fundamental issues remain.

One reason that protein folding remains an unsolved problem is
that the experimental techniques used to study this process do not
provide structural information with atomic detail about the multitude
of folding pathways and intermediates.  Instead the folding process is
generally characterized experimentally by more global properties, such as
the degree of compactness of the protein, the percentage of
secondary structure compared to the native state, different hydrogen
exchange rates of various residues, etc.
The real time observation of protein folding with, for example, NMR,
circular dichroism, or fluorescence techniques, also provides information
about coarse-grained features of the folding intermediates and
pathways with limited spatial resolution.
Consequently, the folding process is frequently analyzed in terms of the
free energy as a function of a heuristic one-dimensional reaction coordinate
which is not directly derived from microscopic motions and thus only reflects
global qualitative features of the folding process.
Computer simulations provide limited information as well because
only short time scales are accessible to realistic simulations.
Furthermore, the large amount of computation required by such simulations
allows the study of only one (or at best a few) trajectories from which
it is difficult to extract general characteristics of the folding process.

The theory presented here potentially lacks many of the limitations
of experiments and computer simulations.
In principle, the theory can predict the time evolution of any
interatomic distance in a polypeptide and can thus specify the average
structure of the polypeptide at any time during the folding or unfolding
process.  However, the practical application of the theory to protein
folding may emerge only after further testing and development.
For example, presently it is unclear how the accuracy of the theoretical
predictions scales with system size.  Therefore, it is desirable to
test the theory for systems larger than pentadecane.
A natural choice would be the helix-coil transition in a simple polyalanine
helix since the computational intensity of nonequilibrium simulations
for a small helix is not significantly greater than that for pentadecane.
Thus, nonequilibrium BD simulations are feasible for the unfolding of the
small helical peptide, thereby again enabling no-parameter tests of the
theory before applying the theory to even larger systems for which the
nonequilibrium simulations are prohibitive.
The recent treatment for the more complicated equilibrium dynamics of
small peptides with the mode coupling theory lends support to
the belief that the nonequilibrium generalization presented here is
adequate for the task of describing the unfolding of a small protein.

A possible direct experimental test of the theory is provided by the
work of Perkins and coworkers.\cite{perkins}  They measure the relaxation
of a single DNA molecule from a highly stretched nonequilibrium state to
equilibrium using optical tweezers to hold the DNA molecule in place
and fluorescence labeling to follow the motion of the DNA molecule.
The relaxation of the end-to-end vector is highly non-exponential, with
a fast initial decay followed by a slower decay.
The simple Rouse and Rouse-Zimm models do not fully explain the
data; this is not surprising since modeling DNA as a string of beads
connected by springs is clearly an oversimplification.
The mode coupling theory provides a framework for treating DNA more
realistically.  Much work remains to be done, however, before a direct
comparison can be made between the experimental data of Perkins and coworkers
and the predictions of the mode coupling theory.  The extension of
the theory to include the solvent would be an important first step, and
the treatment of hydrodynamic interactions is possible following
Ref.~\onlinecite{yi1}
Also, the applicability of the theory to highly charged
polymers, such as DNA, must be tested.  Nevertheless, the predictive power
of the mode coupling theory for describing the dynamics of a variety of
polymers suggests that such generalizations would be fruitful.

\section*{Acknowledgements}
This research is supported, in part, by ACS PRF grants 29067-AC7 and
32263-AC7.
W. T. thanks the Department of Defense for support through an NDSEG
Fellowship.

\bibliography{pentadecane}

\begin{thebibliography}{10}
\bibitem{footnote1}
The symmetry requirements for calculating $\langle R^2_{\rm end} (t)
  \rangle_{\rm noneq}$ are the same as for some of the terms for the
  orientational time correlation function $P_2(t)$, as described in reference
  \onlinecite{tang}.
\end{thebibliography}

\section*{Figure captions}
Fig.~1.  The relaxation of the mean-square end-to-end distance
$\langle R^2_{\rm end}(t) \rangle_{\rm noneq}$
of pentadecane to equilibrium from the all-trans conformation.
The solid line represents the ``exact'' result calculated from the
average of 9000 nonequilibrium BD simulations, while the various
dashed lines represent the results of mode coupling theory calculations
using different basis sets, as described in the text.

Fig.~2.  Same as Fig.~1, except that the
$\langle R^2_{\rm end}(t) \rangle_{\rm noneq}$
axis has a logarithmic scale.

\newpage
\begin{figure}
\psfig{figure=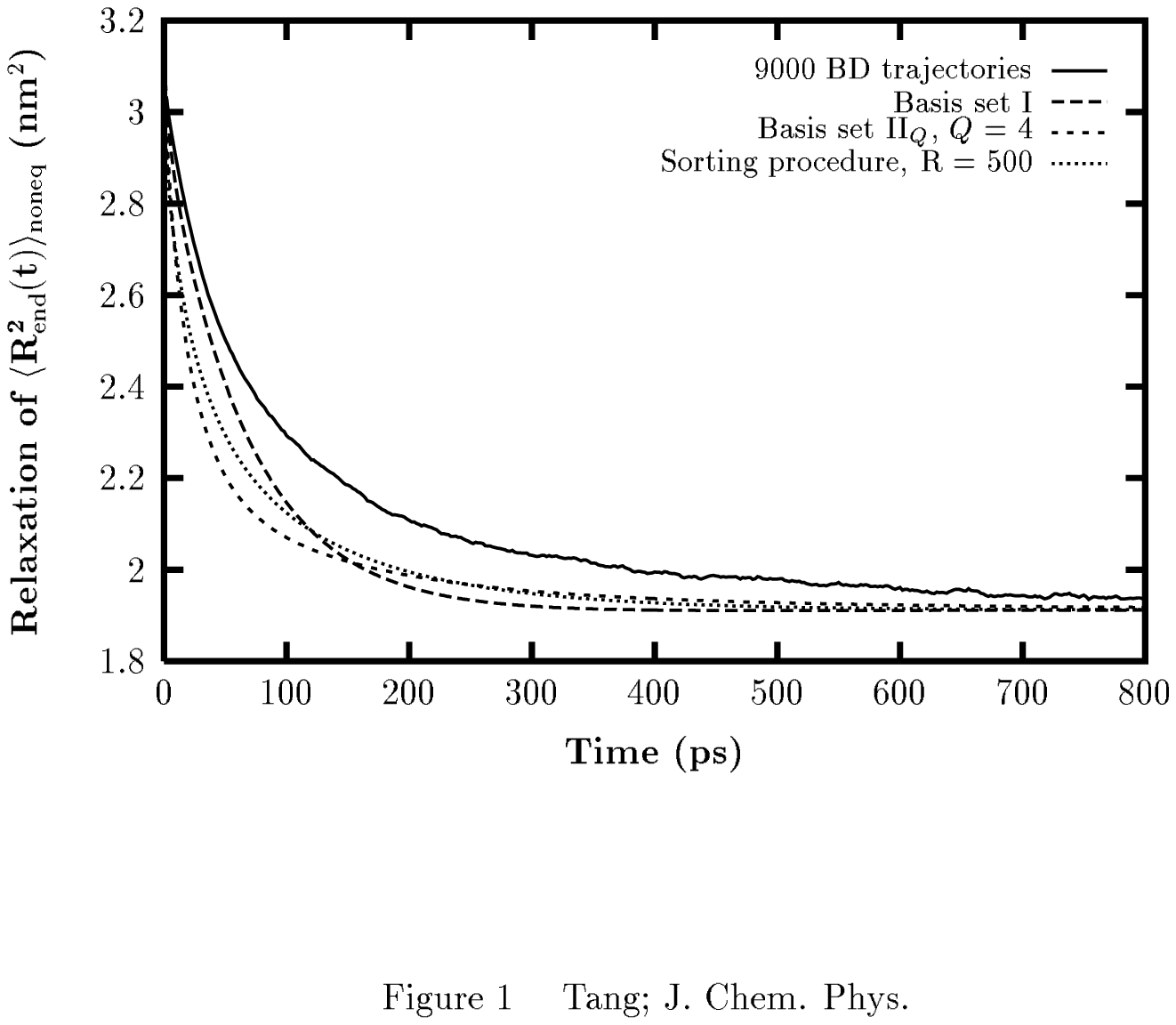}
\end{figure}

\begin{figure}
\psfig{figure=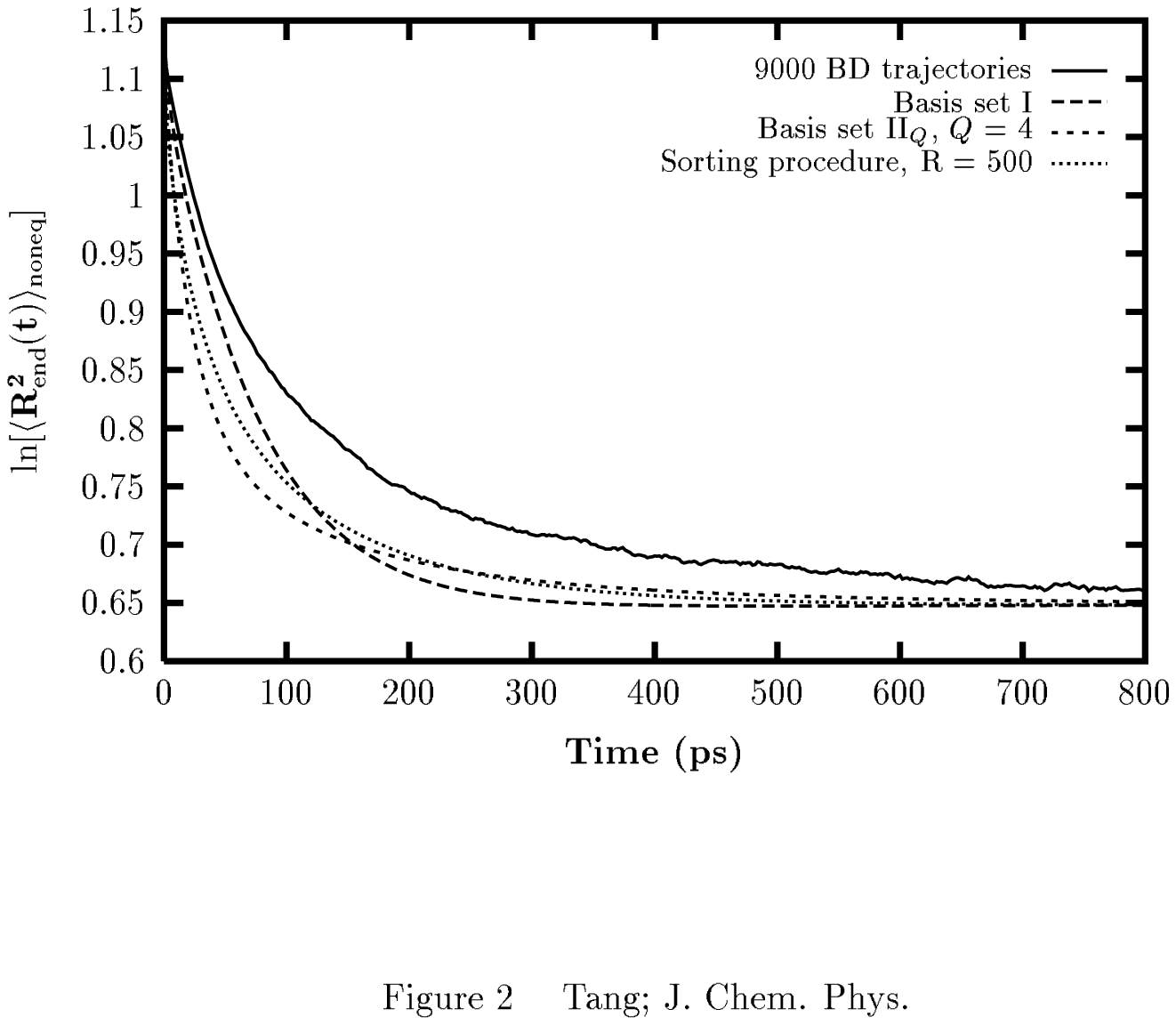}
\end{figure}

\end{document}